\documentclass[aps,prx,twocolumn,showpacs,superscriptaddress,groupedaddress]{revtex4}  
\usepackage{graphicx}  
\usepackage{dcolumn}   
\usepackage{bm}        
\usepackage{amssymb}   

\newcommand{\dd}{\mathrm{d}}
\newcommand{\sigmaz}{\langle\sigma_z\rangle}
\newcommand{\sigmay}{\langle\sigma_y\rangle}
\newcommand{\sigmax}{\langle\sigma_x\rangle}

\begin{document}


\title{Using Spontaneous Emission of a Qubit as a Resource for Feedback Control}
\author{P. Campagne-Ibarcq}

\affiliation{Laboratoire Pierre Aigrain, Ecole normale sup\'erieure, PSL Research University, CNRS, Universit\'e Pierre et Marie Curie, Sorbonne Universit\'es, Universit\'e Paris Diderot, Sorbonne Paris-Cit\'e, 24 rue Lhomond, 75231 Paris Cedex 05, France}
\affiliation{Quantic Team, INRIA Paris-Rocquencourt, Domaine de Voluceau, B.P. 105, 78153 Le Chesnay Cedex, France}

\author{S. Jezouin}

\affiliation{Laboratoire Pierre Aigrain, Ecole normale sup\'erieure, PSL Research University, CNRS, Universit\'e Pierre et Marie Curie, Sorbonne Universit\'es, Universit\'e Paris Diderot, Sorbonne Paris-Cit\'e, 24 rue Lhomond, 75231 Paris Cedex 05, France}
\affiliation{Quantic Team, INRIA Paris-Rocquencourt, Domaine de Voluceau, B.P. 105, 78153 Le Chesnay Cedex, France}

\author{N. Cottet}

\affiliation{Laboratoire Pierre Aigrain, Ecole normale sup\'erieure, PSL Research University, CNRS, Universit\'e Pierre et Marie Curie, Sorbonne Universit\'es, Universit\'e Paris Diderot, Sorbonne Paris-Cit\'e, 24 rue Lhomond, 75231 Paris Cedex 05, France}
\affiliation{Quantic Team, INRIA Paris-Rocquencourt, Domaine de Voluceau, B.P. 105, 78153 Le Chesnay Cedex, France}

\author{P. Six}

\affiliation{Centre Automatique et Syst\`emes, Mines ParisTech, PSL Research University,
60 Boulevard Saint-Michel, 75272 Paris Cedex 6, France.}
\affiliation{Quantic Team, INRIA Paris-Rocquencourt, Domaine de Voluceau, B.P. 105, 78153 Le Chesnay Cedex, France}

\author{L. Bretheau}

\affiliation{Laboratoire Pierre Aigrain, Ecole normale sup\'erieure, PSL Research University, CNRS, Universit\'e Pierre et Marie Curie, Sorbonne Universit\'es, Universit\'e Paris Diderot, Sorbonne Paris-Cit\'e, 24 rue Lhomond, 75231 Paris Cedex 05, France}
\affiliation{Quantic Team, INRIA Paris-Rocquencourt, Domaine de Voluceau, B.P. 105, 78153 Le Chesnay Cedex, France}

\author{F. Mallet}
\affiliation{Laboratoire Pierre Aigrain, Ecole normale sup\'erieure, PSL Research University, CNRS, Universit\'e Pierre et Marie Curie, Sorbonne Universit\'es, Universit\'e Paris Diderot, Sorbonne Paris-Cit\'e, 24 rue Lhomond, 75231 Paris Cedex 05, France}
\affiliation{Quantic Team, INRIA Paris-Rocquencourt, Domaine de Voluceau, B.P. 105, 78153 Le Chesnay Cedex, France}

\author{A. Sarlette}

\affiliation{Quantic Team, INRIA Paris-Rocquencourt, Domaine de Voluceau, B.P. 105, 78153 Le Chesnay Cedex, France}

\author{P. Rouchon}

\affiliation{Centre Automatique et Syst\`emes, Mines ParisTech, PSL Research University,
60 Boulevard Saint-Michel, 75272 Paris Cedex 6, France.}
\affiliation{Quantic Team, INRIA Paris-Rocquencourt, Domaine de Voluceau, B.P. 105, 78153 Le Chesnay Cedex, France}

\author{B. Huard}
\email{benjamin.huard@ens.fr}
\affiliation{Laboratoire Pierre Aigrain, Ecole normale sup\'erieure, PSL Research University, CNRS, Universit\'e Pierre et Marie Curie, Sorbonne Universit\'es, Universit\'e Paris Diderot, Sorbonne Paris-Cit\'e, 24 rue Lhomond, 75231 Paris Cedex 05, France}
\affiliation{Quantic Team, INRIA Paris-Rocquencourt, Domaine de Voluceau, B.P. 105, 78153 Le Chesnay Cedex, France}
\date{\today}

\begin{abstract}
Persistent control of a transmon qubit is performed by a feedback protocol based on continuous heterodyne measurement of its fluorescence. By driving the qubit and cavity with microwave signals whose amplitudes depend linearly on the instantaneous values of the quadratures of the measured fluorescence field, we show that it is possible to stabilize permanently the qubit in any targeted state. Using a Josephson mixer as a phase-preserving amplifier, it was possible to reach a total measurement efficiency $\eta=35~\%$, leading to a maximum of 59~\% of excitation and 44~\% of coherence for the stabilized states. The experiment demonstrates multiple-input multiple-output (MIMO) analog markovian feedback in the quantum regime. 
\end{abstract}

\maketitle

\textit{Introduction}---Decoherence is generally considered as the main limitation to quantum information processing. It can be understood as an exchange of energy and information with the uncontrolled degrees of freedom of the environment leading to vanishing quantum superpositions and relaxation to equilibrium. For a two level system, an ubiquitous source of decoherence comes from spontaneous emission into the electromagnetic modes of the environment. By monitoring the fluorescence of the system, it is then possible to track down its evolution during relaxation~\cite{carmichael1993quantum,wiseman2009quantum,PhysRevX.6.011002,Naghiloo2016}. Here, we describe an experiment that uses the heterodyne detection signal of the fluorescence in order to counteract decoherence and preserve an arbitrary predetermined state of a superconducting qubit. We thus generalize the feedback scheme of previous proposals~\cite{Hofmann1998,Wang2001}, based on the monitoring of a single quadrature of the fluorescence field, to both quadratures so that any predetermined state can be stabilized.
In contrast with previously realized feedback control schemes based on a dispersive quantum non demolition measurement~\cite{Sayrin2011, Vijay2012, Riste2012, Campagne-Ibarcq2013, Riste2013,schindler2013undoing, DeLange2014}, this protocol does not require any extra decoherence or measurement channel in addition to the unavoidable one that is spontaneous emission. Preserving a given qubit state is achieved with finite fidelity by performing rotations around the three axes of the Bloch sphere using driving fields (actuators), whose amplitude depends on the measured quadratures of the fluorescence field (sensors). The experiment thus constitutes a realization, in the most fundamental system, of multiple-input, multiple-output (MIMO) control in the quantum regime~\cite{Chia2011}. This is a key step towards quantum error correction of complex systems~\cite{ahn2003quantum,wang2005dynamical}.

\begin{figure}[!h!t!b!p]
\includegraphics[scale=0.5]{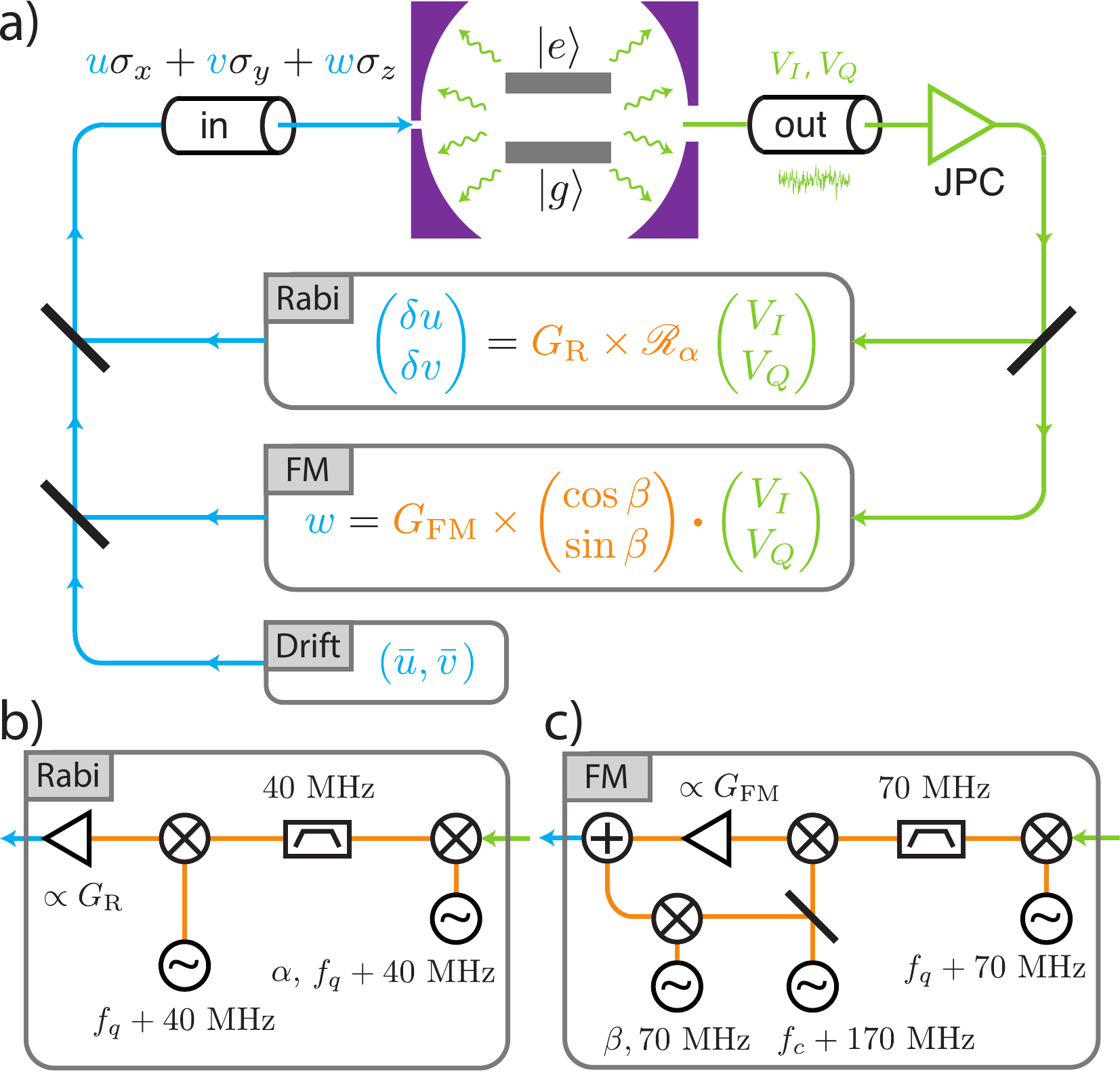}
\caption{\label{fig:schema} \textbf{Feedback scheme and its implementation} (a) An arbitrary state of a transmon qubit is stabilized by feedback control based on continuous  measurement of its fluorescence field quadratures $V_I(t),V_Q(t)$. The control hamiltonian $H_c=\hbar(u(t)\sigma_x+v(t)\sigma_y+w(t)\sigma_z)$ depends linearly on $V_I(t),V_Q(t)$ and is designed to compensate in real-time for the deviations of the qubit from its target state. The controller can be split into three physically distinct boxes. The Drift box adds a constant drive $(\bar u,\bar v)$ for static pre-compensation. The Rabi box adds a drive term $(\delta u,\delta v)$ proportional to the fluorescence field quadratures rotated by an angle $\alpha$. The FM box modulates the qubit frequency leading to a term $w$ proportional to the quadrature $V_\beta=V_I\cos\beta+ V_Q\sin\beta$ of the fluorescence field. (b,c) Scheme of the physical realizations of the Rabi and FM boxes (see text).}
\end{figure}

\begin{figure}[!h!t!b!p]
\includegraphics[scale=0.5]{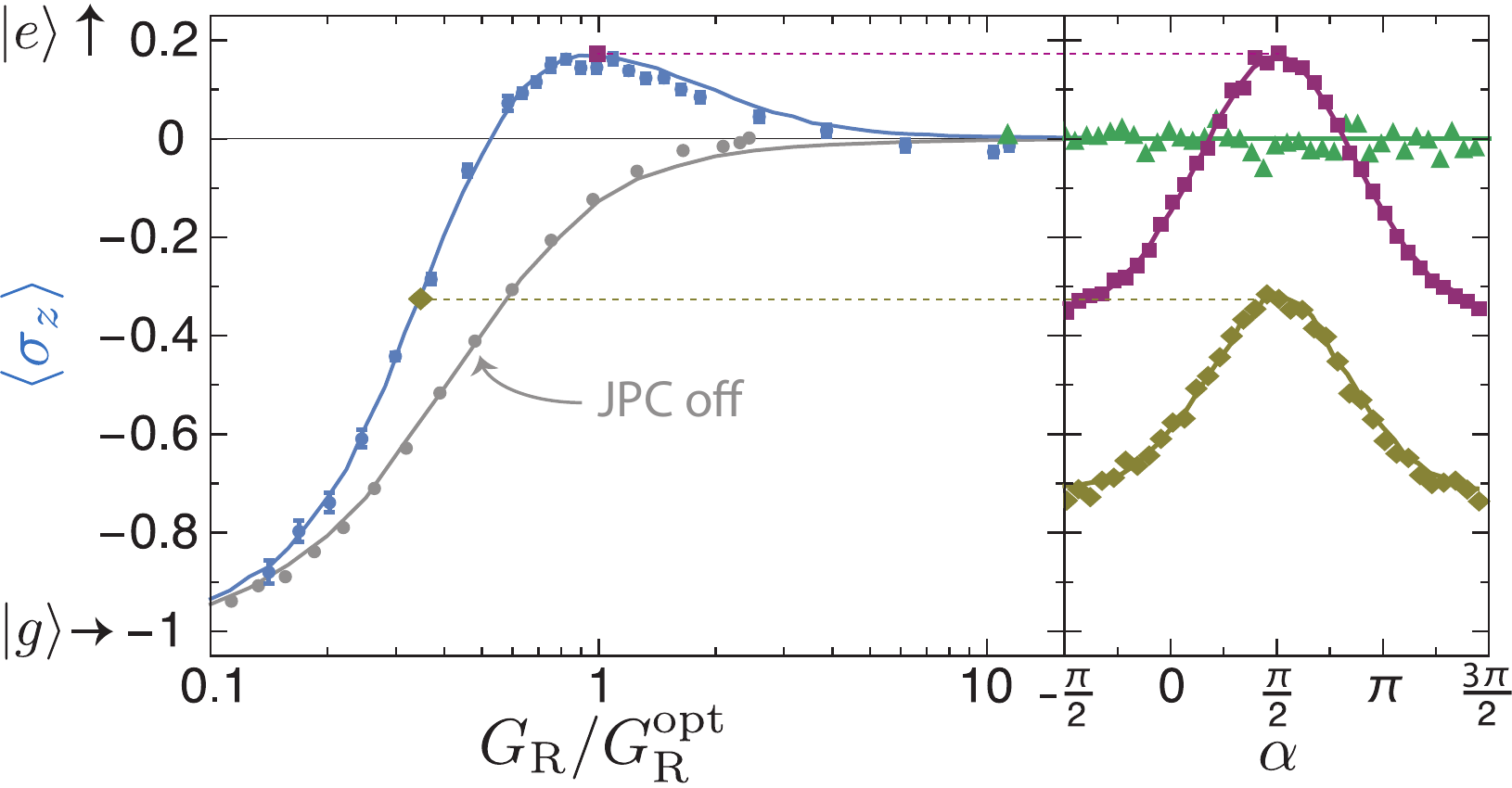}
\caption{\label{fig:excitedstate}  \textbf{Stabilization of the excited state} Targeting the excited state $\vert\Psi_{0,0}\rangle=|e\rangle$, the parameters $G_\mathrm{FM},\bar u,\bar v$ are set to zero according to Eq.~(\ref{eq:LoiFB}) (only "Rabi box" of Fig.~\ref{fig:schema} is on). Measurements (symbols) and simulations (lines) represent $\sigmaz$ as a function of gain $G_R$ for fixed rotation angle $\alpha=\pi/2$ (left panel) and as a function of $\alpha$ for fixed $G_R/G_R^\mathrm{opt}=0.35,1,11.4$ (right panel). The case where the JPC amplifier is turned on (off) is shown in blue (gray). The qubit  is measured after a feedback sequence of 30~$\mu$s ensuring that it is in a steady state. Statistical uncertainties are indicated by error bars in the left panel and by symbols size in the right panel. All the parameters entering the simulations are measured independently.}
\end{figure}

\textit{Feedback protocol}---The fidelity of the stabilization of a quantum state relies crucially on the collection and detection efficiency of the fluorescence field. To maximize collection efficiency, a transmon qubit with transition frequency $f_q=6.27$~GHz is placed inside an off-resonant copper cavity at 30~mK (Fig.~\ref{fig:schema}a). It is designed to channel most spontaneously emitted photons, by Purcell effect~\cite{Purcell46}, through a single dominantly coupled microwave port ("out" in Fig.~\ref{fig:schema}) that accounts for more than 90\% of the total cavity decay rate. The resulting qubit decay rate is measured to be $\gamma_1=(4.7\mathrm{~}\mu\mathrm{s})^{-1}$. The detection efficiency of both quadratures $V_I$ and $V_Q$ of the fluorescence field at $f_q$ is maximized by using a Josephson Parametric Converter (JPC)~\cite{Bergeal2010b, Roch2012,Hatridge2013}. After amplification, the signal is subsequently processed by the analog controller of the feedback loop at room temperature. The two quadratures $V_I$ and $V_Q$ encode information about the qubit relaxation and allow monitoring in real-time of
its quantum trajectory~\cite{PhysRevX.6.011002,Naghiloo2016}. When integrated over a time $\dd t$, they are expressed as
\begin{equation}
\left\{
\begin{array}{r c l}
  V_I\dd t & = & \sqrt\frac{\eta\gamma_1}{2}\langle\sigma_x\rangle \dd t+\dd W_{I}\\
  V_Q\dd t & = & \sqrt\frac{\eta\gamma_1}{2}\langle\sigma_y\rangle \dd t+\dd W_{Q}
\end{array}
\right.,
\label{eq:IQ}
\end{equation}
where $\dd W_{I,Q}$ are independent Wiener processes with variance $\dd t$ modeling zero-point fluctuations and $\sigma_{x,y,z}$ are the Pauli operators. The total measurement efficiency was measured to be  $\eta=35$~\% (see~\cite{supmat}).

The goal of the feedback scheme is to reach and stabilize an arbitrary predetermined qubit state $\vert\Psi_{\theta,\varphi}\rangle=\cos\frac{\theta}{2}\vert e\rangle+\sin\frac{\theta}{2}e^{i\varphi}\vert g\rangle$, where $\theta,\varphi$ parametrize the Bloch sphere. This can be realized by controlling the qubit with the hamiltonian 
$H_c=\hbar(u(t)\sigma_x+v(t)\sigma_y+w(t)\sigma_z)$ that depends linearly on the measured $V_I(t)$ and $V_Q(t)$. Physically, it corresponds to driving the qubit with a microwave signal whose quadratures are modulated as $u(t)=\bar u+\delta u(t)$ and $v(t)=\bar v+\delta v(t)$ and whose angular frequency detuning with the qubit is $w(t)$. The protocol requires three controllers that are schematized by boxes in Fig~\ref{fig:schema}, and whose relevance will become clear in the following sections. 1) The Rabi box consists in driving the qubit with its own fluorescence field that is both amplified by a gain $G_\mathrm{R}$ and phase shifted by $\alpha$. 2) The FM (frequency modulation) box modulates the qubit frequency so that $ w$ is equal to the quadrature $V_\beta= V_I\cos\beta+V_Q\sin\beta$ multiplied by a gain $G_\mathrm{FM}$. 3) The Drift box performs static pre-compensation by adding a constant drive $(\bar u,\bar v)$ independently of $V_{I,Q}$. The parameters of these three controllers determine the stabilized state. We now consider the following set of parameters in order to stabilize $\vert\Psi_{\theta,\varphi}\rangle$,  
\begin{equation}
\begin{array}{l r l r l}
 (1) & G_\mathrm{R} &=\sqrt\frac{\gamma_1}{8\eta}\left(1+\cos\theta\right),  &\alpha &=\pi/2\\
 (2) &  G_\mathrm{FM} &=\sqrt\frac{\gamma_1}{8\eta}\sin\theta, &\beta &=\varphi-\pi/2\\
(3) &   -\frac{\bar u}{\sin\varphi} &= \frac{\bar v}{\cos\varphi} = \frac{\gamma_1}{8\eta}\left(\cos\theta-\eta\right)\sin\theta &&
\end{array}
.
\label{eq:LoiFB}
\end{equation}
This choice is motivated by the ideal case of perfect efficiency ($\eta=1$) and ideal conditions (negligible dephasing, propagation time, and infinite detection bandwidth $B$), where it would stabilize $\vert\Psi_{\theta,\varphi}\rangle$ exactly~\cite{supmat}. As expected, when all boxes are turned off, the qubit is stabilized in the ground state $|g\rangle=|\Psi_{\pi,0}\rangle$.
\begin{figure}[!h!t!b!p]
\includegraphics[scale=0.5]{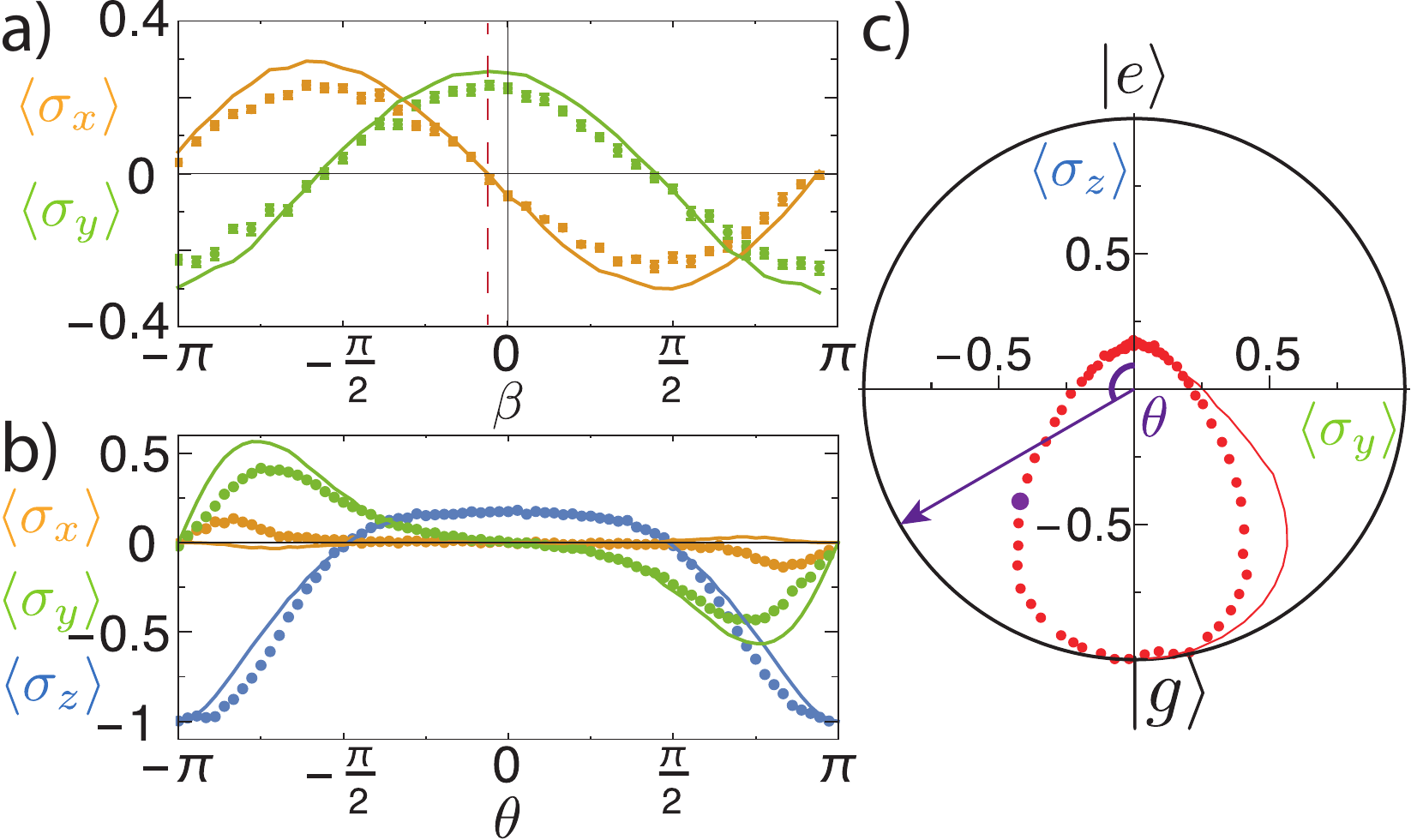}
\caption{\label{fig:ystate}  \textbf{Stabilization of any state} \textbf{a)} Measured (symbols) and simulated (continuous lines) values of $\sigmax$ (orange) and $\sigmay$ (green) as a function of the phase $\beta$ when targeting the state $\vert\Psi_{\pi/2,\pi/2}\rangle=(|e\rangle + i  |g\rangle)/\sqrt{2}$. According to Eq.~(\ref{eq:LoiFB}), one sets $G_R=G_R^\mathrm{opt}/2$, $\alpha=\pi/2$, $\bar u=\gamma_1/8$ and $\bar v=0$. The gain $G_\mathrm{FM}$ is experimentally set to its optimal value $G_\mathrm{FM}^\mathrm{opt}$ giving the maximum $\sigmax^2+\sigmay^2$. The targeted state is most closely reached for $\beta=-10^\circ$ (red) due to a slight non-linearity in the FM box response~\cite{supmat}. As in Fig.~\ref{fig:excitedstate}, comparison with the simulations serves as calibration of the experimental offset on $\beta$. Statistical uncertainties are indicated by error bars (by symbols size in the other panels).
\textbf{b)} Measured (symbols) and simulated (continuous lines) values of $\sigmax$, $\sigmay$ and $\sigmaz$ (respectively orange, green and blue) as a function of the polar angle $\theta$ of targeted state $|\Psi_{\theta,\pi/2}\rangle$. \textbf{c)} Projection of the Bloch sphere on the $yz$-plane. Red dots represent the measured stabilized states of (b). The corresponding simulated states are shown as a red line. A purple dot indicates the stabilized states when aiming for $\theta=2\pi/3$.}
\end{figure}

\textit{Stabilizing the excited state}---To start with, we aim at stabilizing the excited state $\vert\Psi_{0,0}\rangle=\vert e\rangle$ of the qubit. This target only involves the Rabi box of the feedback loop since, according to Eq.~(\ref{eq:LoiFB}), $G_\mathrm{FM}=\bar u=\bar v=0$ when $\theta=0$.  The fluorescence field is phase shifted by $\alpha=\pi/2$ before being sent back to the qubit. This can be qualitatively understood by noting that $V_I$ (respectively $V_Q$) gives information about qubit deviations from $\vert e\rangle$ in the $x(y)$-direction of the Bloch sphere, which can be compensated for by applying rotations around the orthogonal $y(x)$-direction.



To implement the Rabi box (Fig.~\ref{fig:schema}b), the fluorescence field is first downconverted to 40 MHz and phase shifted by a mixer whose local oscillator has a tunable phase offset that is able to set $\alpha$~\cite{supmat}. 
Then, this 40~MHz signal is filtered in the band 25-50~MHz in order to avoid heating the qubit with an important noise power, particularly at the transition frequency towards the higher energy levels of the transmon. Finally, the signal is upconverted back to $f_q$ 
and a series of amplifiers and variable attenuators allows one to control the overall gain $G_R$ of the Rabi box.


By varying the parameters $G_R$ and $\alpha$, it is possible to maximize the excitation of the stabilized state (Fig.~\ref{fig:excitedstate}). The excitation is characterized by measuring the qubit after turning on the feedback protocol during a time $30~\mu\mathrm{s}$ (much longer than $\gamma_1^{-1}$). Note that all shown tomographic measurements in this letter are obtained by averaging the fluorescence signal (feedback off) over $5~\mu\mathrm{s}$ with or without an initial $\pi/2$ pulse on $\sigma_x$~\cite{Houck2007,supmat}. 
When setting $\alpha=\pi/2$, $\sigmaz$ is measured as a function of the gain $G_R$, and exhibits a maximum $\sigmaz=0.17$ at $G_R^{opt}$ corresponding to an inverted population with 59~\% of excitation (blue in left panel of Fig.~\ref{fig:excitedstate}). For the largest gains, the qubit reaches a maximal entropy state as evidenced by the measured $\sigmaz=0$ whatever $\alpha$ (green in right panel of Fig.~\ref{fig:excitedstate}). There, control signals produce so large and noisy rotation angles that the qubit state is effectively randomized. Note that the coherences $\sigmax$ and $\sigmay$ were measured to be zero for all gain $G_R$. When the JPC is turned off ($\eta\simeq 0.005$, grey symbols), we observe that $\sigmaz$ increases monotonically with $G_R$ but saturates at $0$. There, the detection efficiency is so small that the inputs of the Rabi box are just noisy signals with negligible correlation with the qubit state. The qubit is then effectively heated by a thermal source and population inversion never occurs.

The stationary state in presence of feedback can be computed by solving numerically a stochastic master equation~\cite{wiseman1993quantum,wiseman1994quantum,Chia2011,supmat}.  The relevant parameters, which are all measured independently, are the decay rate $\gamma_1$, measurement efficiency $\eta$, dephasing rate $\gamma_\phi=(22\mathrm{~}\mu\mathrm{s})^{-1}$, finite bandwidth of the detection setup $B=3.3$~MHz and non-zero delay time $T_d\approx 0.12~\mu\mathrm{s}$.
The data in both panels of Fig.~\ref{fig:excitedstate} are in excellent agreement with simulations of the stochastic master equation (continuous lines). Further simulations show that the stabilized excitation is mainly limited to 59~\% because of the measurement inefficiency (see Table 1 in ~\cite{supmat}). Note that in the experiment the gain $G_R$ (resp. phase $\alpha$) is known up to some prefactor (resp. offset). The simulations, which show that the maximum occurs at $G_R^\mathrm{opt}=\sqrt{\gamma_1/2\eta}$ ($\alpha=\pi/2$) allow to calibrate it.

\textit{Stabilizing a coherent superposition}---
We now turn to stabilizing an arbitrary coherent state. In the work of Wang and Wiseman~\cite{Wang2001}, it is shown that rotations around $\sigma_x$ and $\sigma_y$ alone (Rabi box) cannot stabilize states on the equator of the Bloch sphere. Therefore we implement a feedback protocol that also uses rotations around the $\sigma_z$ axis, which corresponds to modulating the qubit frequency (FM box in Fig.~\ref{fig:schema}c). It relies on AC Stark shift effect: a drive close to cavity frequency $f_c=7.86$~GHz induces an intensity dependent shift of the qubit frequency~\cite{Gambetta2006,Blais2007}. For that purpose, the fluorescence signal is bandpass filtered (bandwidth $B_f=2~\mathrm{MHz}<B$) and upconverted 
to a frequency $f_c+\Delta$. We use a large enough detuning $\Delta=100$~MHz 
so that the measurement induced dephasing rate $\gamma_m=(84\mathrm{~}\mu\mathrm{s})^{-1}$ remains much smaller than $\gamma_1$ 
for the parameters required by the feedback law~\cite{supmat}. In order to get $w$ proportional to a single quadrature $V_\beta$, we combine the upconverted field, amplified by a gain $G$, with a large amplitude $\epsilon_0e^{i2\pi(f_c+\Delta)t+i\beta}$ tone. Then to first order,
%
%
\begin{equation}
w(t)/2\pi \propto \vert \epsilon_0e^{i\beta}+G(V_I+iV_Q)\vert^2 \simeq  \epsilon_0^2+2G\epsilon_0V_\beta. \label{EqwofA0}
\end{equation}
The gain $G_\mathrm{FM}\propto G\epsilon_0$ of the FM box can thus be tuned by varying $G$~\cite{supmat}. In practice, the frequency offset due to $\epsilon_0^2$ is equal to 810~kHz and simply renormalizes $f_q$ throughout all experiments. 

To begin with, one can stabilize the state $\vert\Psi_{\pi/2,\pi/2}\rangle=(|e\rangle+i|g\rangle)/\sqrt{2}$ on the equator of the Bloch sphere. Therefore according to Eq.~(\ref{eq:LoiFB}), we set $\alpha=\pi/2$, $G_R=G_R^\mathrm{opt}/2$, $\bar{u}=\gamma_1/8$ and $\bar{v}=0$. The gain $G_\mathrm{FM}$ is then empirically set to the value $G_\mathrm{FM}^\mathrm{opt}$ that maximizes the coherences $\sqrt{\sigmax^2+\sigmay^2}$.  The optimal value for $\beta$ can be found in Fig.~\ref{fig:ystate}a, which shows the measured coherences $\sigmax$ and $\sigmay$ of the state stabilized by the feedback loop as a function of $\beta$. Simulations of the stochastic master equation are shown as continuous lines. 

As can be seen, for $\beta=-10^\circ$, the Bloch vector is in the direction of the target state with $\sigmax=0$ and $\sigmay\approx 0.22$. The slight offset in $\beta$ from 0 compared to Eq.~(\ref{eq:LoiFB}) originates~\cite{supmat} from the neglected higher order terms in Eq.~(\ref{EqwofA0}). Now varying $\beta$, it can be seen (Fig.~\ref{fig:ystate}a) that $\sigmax$ and $\sigmay$ follow sinusoidal oscillations in quadrature. It indicates that any state on the equator of the Bloch sphere can be stabilized by just changing $\beta$, even though, here, the drifts $\bar u$ and $\bar v$ have not been adjusted to their optimal values for each value of $\beta$.  
We notice an overestimation of the amplitude of the coherences by the simulations. This mismatch between experiment and simulation for $\theta>0$ remains when considering various parasitic effects, such as pollution of $I$ and $Q$ by the transmitted control fields through the cavity, or the impact of internal cavity dynamics on measurement induced dephasing and real-time qubit frequency modulation~\cite{supmat}.
\begin{figure}[!h!t!b!p]
\includegraphics[scale=0.55]{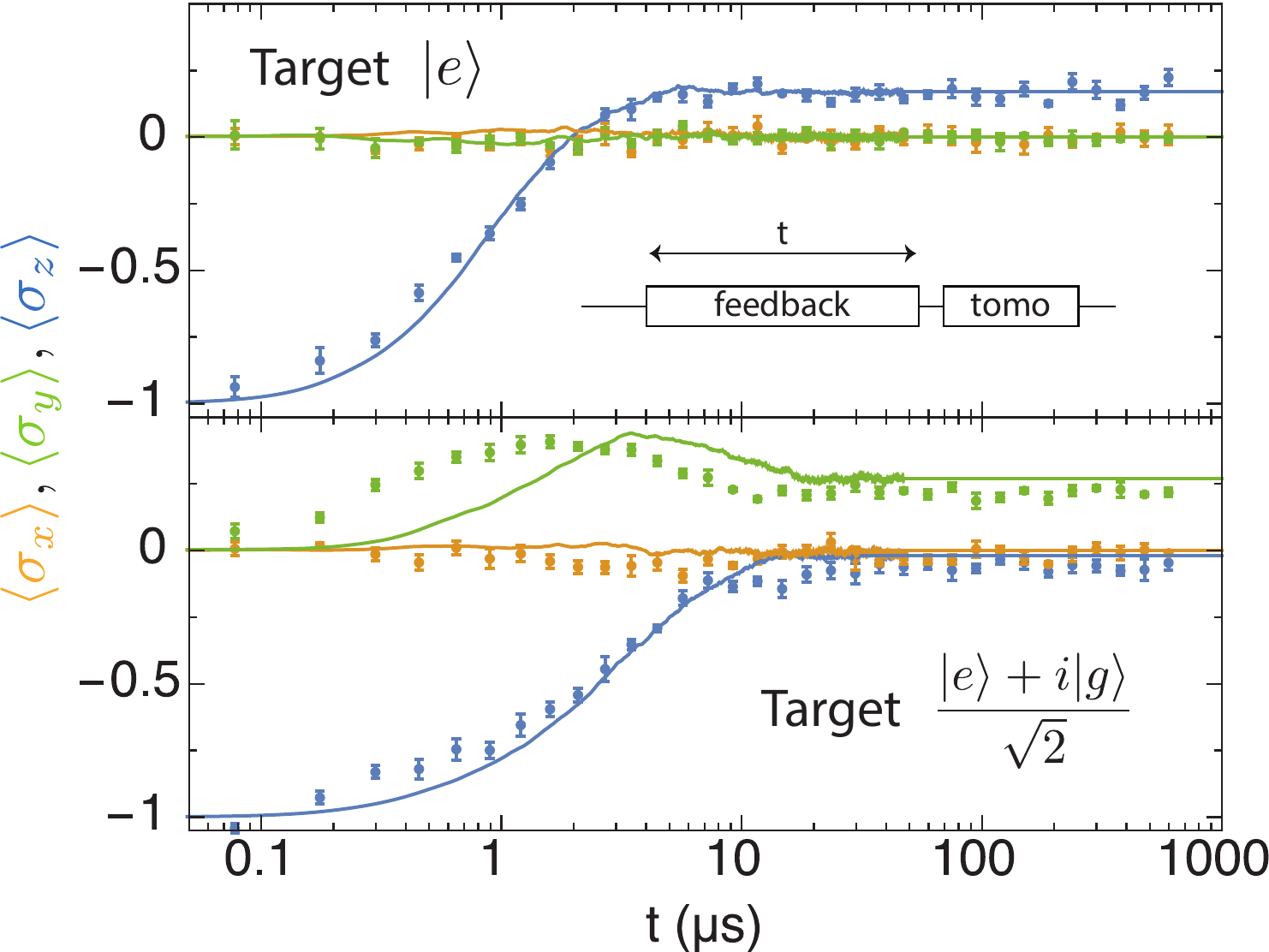}
\caption{\label{fig:time} \textbf{Transient dynamics under feedback} Measured (symbols) and simulated (continuous lines) $\sigmax$, $\sigmay$ and $\sigmaz$ (respectively orange, green and blue) as a function of the duration $t$ of the feedback control. At $t=0$, the qubit is in thermal equilibrium and all control parameters in Fig.~\ref{fig:schema} are set to $0$. The upper panel corresponds to the target state $\vert\Psi_{0,0}\rangle=\vert e\rangle$ ($G_R=G_R^\mathrm{opt}$, $\alpha=\pi/2$ as for the purple in Fig.~\ref{fig:excitedstate}left). The lower panel corresponds to the target state $\vert\Psi_{\pi/2,\pi/2}\rangle=\left(\vert e\rangle+i\vert g\rangle\right)/\sqrt{2}$ ($\beta=-10^\circ$ as for the red in Fig.~\ref{fig:ystate}a). Error bars are statistical uncertainties. Simulations are performed with the same model as in Figs.~\ref{fig:excitedstate},\ref{fig:ystate}a}
\end{figure}

Since we have already shown how to tune the longitude of the stabilized state (Fig.~\ref{fig:ystate}a), we now fix $\beta=-10^\circ$ and proceed to stabilize states $|\Psi_{\theta,\pi/2}\rangle$ of arbitrary polar angle $\theta$. Measured stationary values of $\sigmax,\sigmay$ and $\sigmaz$ are plotted in Fig.~\ref{fig:ystate}b as a function of $\theta$ and in the Bloch sphere representation in Fig.~\ref{fig:ystate}c. The gain of the FM box is here set to $G_\mathrm{FM}= G_\mathrm{FM}^\mathrm{opt}\sin\theta$ and the other feedback parameters obey Eq.~(\ref{eq:LoiFB}). 
As expected from feedback correcting for relaxation, the purity of the stabilized state is larger for targeted states closer to $|g\rangle$ (smaller values of $\sigmaz$). The largest stabilized coherences reach 44~\% and the feedback efficiency~\cite{Vijay2012} is 0.43 on average. The latter are mainly limited by efficiency $\eta$ and dephasing (see Table 1 in ~\cite{supmat}). 
The mismatch between simulations and measurements is even stronger when targeting states with $\sigmaz<0$ and measured purities are significantly below expectation. The emergence of a non-zero $\sigmax$ close to the ground state in Fig.~\ref{fig:ystate}b can be explained by a constant detuning of the control fields $\bar{u}$ and $\bar{v}$ of a few kHz. Note that the feedback law Eq.~(\ref{eq:LoiFB}) does not maximize the purity when $\eta<1$. We propose an optimal scheme in absence of dephasing and loop delay in~\cite{supmat}.

\textit{Transient behavior}--- In other continuous Markovian feedback schemes based on a dispersive measurement of a cavity~\cite{Vijay2012} (respectively on fluorescence~\cite{Hofmann1998,Wang2001}), the convergence rate towards the target state decreases towards zero as the target state approaches $|g\rangle$ or $|e\rangle$ (resp. $|\Psi_{\pi/2,\varphi}\rangle$). In contrast, we show in~\cite{supmat} that, for any target on the Bloch sphere and any value of $\eta$, the qubit converges to the closest possible state at least at an exponential rate $1/T_1$. By varying the duration of the feedback protocol, it is possible to determine its dynamics when the qubit is initially in $|g\rangle$ (Fig.~\ref{fig:time}). When targeting the excited state we observe, in excellent agreement with the simulations, that $\sigmaz$ rises exponentially with a rate of about $4\gamma_1$. Once the stationary state is reached, it remains there permanently. When targeting the equator $\sigmaz$ still exhibits an exponential increase at a rate of about $1.5\gamma_1$ towards a stationary value, which is here just below zero due to imperfections of the feedback loop. In contrast, $\sigmay$ shows a bump at small times, showing that coherences initially increase for a few $\mu\mathrm{s}$ and then decrease. This behavior is qualitatively predicted by the simulations although we do not obtain quantitative agreement when targeting the equator possibly for the same reason as in Fig.~\ref{fig:ystate}a.
%

\textit{Conclusion}---
This set of experiments thus demonstrates that the information contained in both quadratures of the fluorescence of a qubit can be used to preserve an arbitrary predetermined state with finite fidelity using analog Markovian feedback. This technique is not exclusive but complementary to QND measurement based feedback~\cite{Sayrin2011, Vijay2012, Riste2012, Campagne-Ibarcq2013, Riste2013,schindler2013undoing, DeLange2014}.
More generally, Markovian feedback can effectively modify the backaction associated to a measurement, which amounts to effectively engineering dissipation~\cite{schirmer2010stabilizing}, in a similar manner to reservoir engineering schemes. The stabilization of manifolds in Hilbert spaces with more dimensions than two, which was recently demonstrated using autonomous feedback in qubit registers~\cite{shankar2013autonomously,lin2013dissipative} or harmonic oscillators~\cite{leghtas2015confining}, may have counterparts using continuous measurement  feedback~\cite{ahn2003quantum,wang2005dynamical} that could prove to be useful to quantum error correction. Our results could be applied to the wide variety of physical systems that decay by fluorescence.


\textbf{Acknowledgements} We thank Michel Devoret, Emmanuel Flurin, Vladimir Manucharyan, Mazyar Mirrahimi and the Quantronics group for fruitful discussions. Nanofabrication has been made within the consortium Salle Blanche Paris Centre. This work was supported by the EMERGENCES grant QUMOTEL of Ville de Paris and by the IDEX program ANR-10-IDEX-0001-02 PSL$^*$.
P.C.I. and S.J. contributed equally to this work.   

\bibliographystyle{landry}

\end{document}